# Solution to the Equity Premium Puzzle Using the Sufficiency

# Factor of the Model

Atilla Aras

**Declarations of interest**: none.

**Funding**: This research did not receive any specific grant from funding agencies in the public, commercial, or not-for-profit sectors.

# Solution to the Equity Premium Puzzle Using the Sufficiency

# Factor of the Model

## Abstract

This study provides the solution to the equity premium puzzle. The new model was developed by including the behavior of investors toward risk in financial markets in prior studies. The calculations of this newly tested model show that the value of the coefficient of relative risk aversion is 1.033526 by assuming the value of the subjective time discount factor to be 0.99. Since these values are compatible with the existing empirical studies, they confirm the validity of the newly derived model that provides the solution to the equity premium puzzle.

## 1. Introduction

The equity premium puzzle is a quantitative puzzle that implies the inability of intertemporal economic models to explain the large historical equity premium under reasonable parameter values in US financial markets over the past century. The equity premium puzzle, coined by Mehra and Prescott (1985), arises because a high historical equity premium leads to an unreasonably high level of risk aversion among investors according to standard intertemporal economic models. The puzzle is important for both the academic world and practitioners because intertemporal economic models do not replicate such a high historical equity premium under reasonable parameters. In this study, the solution for this unsolved puzzle is proposed.

The motivation of the study stems from the implicit assumption of the fair gamble that the behavior of individuals toward risk can be expressed by means of utility curves. The study is mainly a modification of prior studies by observing a variable that was unknown until now.

Individuals decide about an uncertain value by comparing its utility with that of a certain value. The behavior of individuals toward risk, as risk-averse, risk-loving, or risk-neutral decision-makers, can be observed at the time the individuals compare the utilities of certain and uncertain wealth values. The risk-taking or risk-averse behavior of investors occurs after they allocate additional positive or negative utility for the uncertain wealth value at the time of this comparison. Hence, definitions of risk-averse, risk-loving or risk-neutral investors have been reformulated in this study.

The problem of a typical investor has been reformulated in the new model. Following Mehra (2008), Cochrane (2009), and Danthine and Donaldson (2014), a new system of equations has been developed.

The values of the coefficient of relative risk aversion and the sufficiency factor of the model for investors who invested in equity and risk-free assets in the US economy for the 1889-1978 period were calculated using the new system of equations.

The coefficient of relative risk aversion and the subjective time discount factor are 1.033526 and 0.99, respectively. Since these values are compatible with empirical studies, they confirm that the new model gives a solution to the equity premium puzzle.

## 2. Materials and methods

The standard intertemporal economic models do not explain that the large historical equity premium has led to an implausibly high level of risk aversion in US financial markets in the past century. This is the equity premium puzzle that was coined by Mehra and Prescott (1985).

Mehra and Prescott coined the equity premium puzzle by arguing that the consumption capital asset pricing model (CCAPM) is unable to explain the high historical US equity premium over the past century under reasonable parameter values. The CCAPM is an economic model that provides a theoretical connection between aggregate consumption and financial markets (Danthine & Donaldson, 2014). The CCAPM indicates that the representative agent's utility function, his or her subjective time discount factor, and the fact that consumption equals dividends in the exchange economy equilibrium are the only factors that determine security returns (Danthine & Donaldson, 2014). To date, no agreed-upon solution for the puzzle has been provided, which means that the CCAPM replicates this high equity premium under reasonable parameter values.

The studies that propose solutions to the equity premium puzzle can be grouped into seven classes (Mehra 2003; Mehra 2008).

The first class consists of existing literature that presents preference modifications for the solution (Abel, 1990; Benartzi & Thaler, 1995; Campbell & Cochrane, 1999; Costantinides, 1990; Epstein & Zin, 1991).

The second class consists of studies that present survival bias for the solution (Brown et al., 1995).

The third class is based on studies that present market imperfections for the solution (Aiyagari & Gertler, 1991; Alvarez & Jermann, 2000; Bansal & Coleman, 1996; Costantinides et al., 2002; Heaton & Lucas, 1996; McGrattan & Prescott, 2001).

The fourth class consists of studies that present modified probability distributions to admit rare but disastrous events for the solution (Rietz, 1988).

The fifth class is based on existing literature that presents incomplete markets for the solution (Costantinides & Duffie, 1996; Heaton & Lucas, 1997; Mankiw, 1986).

The sixth class comprises studies that present limited participation of consumers in the stock market for the solution (Attanasio et al., 2002; Brav et al., 2002).

The seventh class consists of studies that present temporal aggregation problems for the solution (Gabaix & Laibson, 2001; Heaton, 1995; Lynch, 1996).

Individuals choose between certain and uncertain wealth values by comparing their utilities. None of the proposed solutions consider that individuals may increase or decrease their utility of uncertain wealth value according to their risk preferences at the time of the abovementioned comparison. This observation of the author of the article will be incorporated into the new model in subsequent sections.

### 3. Theory

## 3.1 Some Issues in the Standard Definitions of the Risk Behavior of Investors

Certain utility curves of wealth values are assumed to be different for risk-averse, risk-loving and risk-neutral investors in the standard definitions of the risk behavior of investors. To confirm these definitions, the concept of the risk premium is shown as a proof. $\pi \cong \frac{1}{2} \sigma^2 \alpha(w)$ is the risk premium that reflects the amount the investor must pay to avoid the gamble. Here, $\sigma^2$ is the variance of the gamble, and $\alpha(w)$ is the coefficient of absolute risk aversion. Risk premiums must be positive for risk-averse investors and must be negative for risk-loving investors. The risk premium is positive when the certain utility curve of wealth is concave and negative when the certain utility curve of wealth is convex.

There is no mention of risk for events whose outcome is certain. In other words, there is no chance of risk-taking or risk-avoidance for events whose outcome is certain. Since certain utility curves imply the utilities of certain outcomes, certain utility curves cannot be different for risk-averse and risk-loving investors as they are different in the standard definitions of the risk behavior of investors.

The comparison of the utilities of certain and uncertain wealth values is conducted at the same wealth value, and the subjective time discount factor for the utility of an uncertain wealth value is not used in the standard definitions of the risk behavior of investors. However, there is no need to compare the utilities of certain and uncertain wealth values at the same wealth value. For instance, the attitude of risk-averse investors toward risk at the time of a comparison may lead to a reduction in the utility of the uncertain wealth value below the utility of the certain wealth value that is of a different magnitude from the utility of the uncertain wealth value. This comparison leads investors to choose a certain wealth value. Therefore, the behavior of risk-averse individuals is such that the negative utility allocated for the uncertain wealth level at the time of a comparison reduces the utility of the uncertain wealth level below the utility of the certain wealth level. Hence, the comparison should be performed between the utilities of certain and uncertain values that may be at different magnitude levels, but the utility of an uncertain value must be discounted by the subjective time discount factor. For instance, a certain utility of $25 can be compared with the predicted uncertain utility of $30 after including the subjective time discount factor and negative utility allocation for the uncertain wealth level for a risk-averse investor.

## 3.2 Sufficiency Factor of the Model

In finance theory, when investors make a decision about an uncertain wealth value, the standard method is to design a lottery and compare the expected value of its utility to the utility of a certain wealth value. The expected value of the lottery is estimated to determine the payoff of the lottery that is a future variable. However, the payoff of the lottery determined by this method may be incorrect due to future uncertainty or the insufficiency of the method used. Since it is impossible to predict future values correctly most of the time using any method, investors will not trust that method completely. Therefore, it is necessary to reflect future uncertainty and the insufficiency of the method used on the uncertain utility gained from the future values at the time individuals compare the utilities of certain and uncertain values. Risk-taking or risk-averse behavior of investors occurs after they allocate additional positive utility or negative utility for an uncertain wealth value due to the insufficient model used and future uncertainties. Thus, an investor may increase or decrease his uncertain utility by considering this future uncertainty and insufficiency of the method used according to his or her risk preferences. Hence, a related parameter must be reflected in the uncertain utility at the time of this comparison. Investors allocate positive, negative or zero utility for uncertain wealth levels; and then, according to the nature of this allocation, they are classified as risk averse, risk loving or risk neutral. This reflection on uncertain utility shows us the behavior of individuals toward risk as risk-averse, risk-loving, or risk-neutral decision-makers.

SOLUTION TO THE EQUITY PREMIUM PUZZLE

Hence, the definitions for risk-averse, risk-loving, and risk-neutral investors are reformulated in this study as follows.

Uncertain utility is assumed to be predicted by the conditional expectation operator. The conditional and unconditional expectations are assumed to be the same. Since there is no chance of avoiding or taking risk for individuals in situations whose outcomes are certain, certain utility curves are assumed to be the same (i.e., concave) for all types of investors. The drawback of this assumption is that risk premiums cannot be derived for risk-loving investors using the existing standard techniques.

*A risk-averse investor* allocates negative or zero utility to the uncertain wealth value due to the insufficient model used and future uncertainties at time t such that

$$u(w_t) > \beta \eta_t E_t[u(w_{t+1})] \tag{1}$$

holds true. Here, u is a continuously differentiable and increasing concave utility curve, t is the time the investor compares the utilities of certain and uncertain wealth values, $u(w_t)$ is the certain utility of a wealth value at time t, $E_t[u(w_{t+1})]$ is the predicted uncertain utility gained from future wealth value $(w_{t+1})$ with the information set available at time t, $\beta$ is the subjective time discount factor and $\eta_t$ is *the sufficiency factor of the model* that is coined by the author of the article. $\eta_t$ is a coefficient that is determined at time t for the utility curve of the uncertain value, that is, $E_t[u(w_{t+1})]$. It is calculated as follows:

$\eta_t E_t[u(w_{t+1})] = E_t[u(w_{t+1})]$ + negative utility allocated by the investor at time t due to the insufficient model used and future uncertainties,

$\eta_t E_t[u(w_{t+1})] = E_t[u(w_{t+1})]$ + positive utility allocated by the investor at time t due to the insufficient model used and future uncertainties, and

$\eta_t E_t[u(w_{t+1})] = E_t[u(w_{t+1})]$ + zero utility allocated by the investor at time t due to the insufficient model used and future uncertainties.

$\eta_t$ is determined at time t because positive, negative or zero utility due to the insufficient model used and future uncertainties are allocated by the investor for the uncertain wealth value at time t. Time t denotes the time the investor compares the certain and uncertain utility values. Hence, risk-averse, risk-loving or risk-neutral behavior is an activity at time t, that is, risk behavior is observed at time t.

*A risk-loving investor* allocates positive utility for the uncertain wealth value due to the insufficient model used and future uncertainties at time t so that

$$u(w_t) < \beta \eta_t E_t[u(w_{t+1})] \tag{2}$$

holds true. This is because there is a chance that the investor may profit from the insufficient model and future uncertainties.

A *not enough risk-loving investor*, however, allocates positive utility for the uncertain wealth value due to the insufficient model used and future uncertainties at time t so that

$$u(w_t) > \beta \eta_t E_t[u(w_{t+1})] \tag{3}$$

holds true.

Finally, a *risk-neutral investor* allocates positive utility for the uncertain wealth value due to the insufficient model used and future uncertainties so that

$$u(w_t) = \beta \eta_t E_t[u(w_{t+1})] \tag{4}$$

holds true.

An example will be given to make the definitions more concrete. An investor will decide to buy or not buy a financial asset in the current period t by considering the following inequalities. Assume that a risk-averse investor is planning to buy a stock that has a price of $2 in current period t. Planning to receive a dividend equal to $0.1, the investor expects the price of the stock to rise to \$2.1 in the next period. Because he or she is risk averse, he or she allocates negative utility for the uncertain price of the stock so that

$u(\$2) > \{\beta E_t[u(\$2.2)]\} +$ {negative utility allocated by the investor in the current period t}

$= u(\$2) > \beta \eta_t E_t[u(\$2.2)]$ holds true.

Hence, the investor will decide to buy the financial asset in current period t.

The certain and uncertain utility curves of a risk-averse investor after including the sufficiency factor of the model and the subjective time discount factor are demonstrated in Figure 1. [Insert Figure 1 here]

The relationships between certain and uncertain utility curves considering the sufficiency factor of the model and the coefficient of relative risk aversion (ρ) will be as follows:

$\eta u(w)$ will be below $u(w)$ if

$$\eta < 1 \text{ and } \rho < 1 \qquad (5)$$

and

$$\eta > 1 \text{ and } \rho > 1 \qquad (6)$$

hold true.

$\eta u(w)$ will be above $u(w)$ if

$$\eta > 1 \text{ and } \rho < 1 \qquad (7)$$

and

$$\eta < 1 \text{ and } \rho > 1 \qquad (8)$$

SOLUTION TO THE EQUITY PREMIUM PUZZLE

hold true.

The above inequalities are demonstrated in Figures 2 and 3. [Insert Figure 2 here]

[Insert Figure 3 here]

### 3.3 Solution to the Equity Premium Puzzle with the Sufficiency Factor of the Model

In this section, a new model has been developed to solve the equity premium puzzle after modifying the following equations from prior studies:

$$max_{\{\theta\}}\, u(\,c_t) + E_t[\beta u(c_{t+1})]$$

$$\text{s.t.} \tag{9}$$

$$c_t = e_t - p_t\theta$$

$$c_{t+1} = e_{t+1} + d\theta$$

(Cochrane, 2009, p. 5),

$$p_t u'(c_t) = \beta E_t[(p_{t+1} + y_{t+1})u'(c_{t+1})] \tag{10}$$

(fundamental pricing relationship) (Cochrane, 2009, p. 5),

$$ln\, E_t\,(R_{e,t+1}) = -\,ln\,\beta + \rho\mu_x - \frac{1}{2}\rho^2{\sigma_x}^2 + \rho\sigma_{x,z} \tag{11}$$

(Mehra, 2008, p. 19),

$$ln\, R_f = -\,ln\,\beta + \rho\mu_x - \frac{1}{2}\rho^2{\sigma_x}^2 \tag{12}$$

(Mehra, 2008, p. 19),

$$ln\, E(R_e) - ln\, R_f = \rho{\sigma_x}^2 \tag{13}$$

SOLUTION TO THE EQUITY PREMIUM PUZZLE

(Mehra, 2008, p. 19), and

$$E(R_e) = \frac{E(x_{t+1})}{\beta E(x_{t+1}^{1-\rho})} \qquad (14)$$

(Danthine & Donaldson, 2014, p. 291).

### 3.3.1 New Model

The behavior of investors toward risk in financial markets will be included in prior studies through the sufficiency factor of the model. Investors cannot predict future prices correctly most of the time. Hence, they allocate additional positive, negative or zero utility for future prices. These additional allocations by the investors are achieved through the sufficiency factor of the model. Investors are then classified as risk-averse, risk-loving and risk-neutral decision-makers according to the nature of this allocation. Hence, a solution to the equity premium puzzle will be given by including the sufficiency factor of the model in the typical problem of the investor.

The theoretical formulas of prior studies involve conditional expectations, conditional variances, and conditional correlations of consumption growth and returns. However, these conditional moments are replaced by sample means, variances, and correlations in standard empirical tests (Munk, 2013, p.294). These empirical tests assume that all of the observations of the consumption growth rate and returns are drawn from the same distribution with a constant conditional expectation and a conditional variance (Munk, 2013, p.294). Hence, it can be concluded that the traditional empirical tests are able to test the constant moments, the constant relative risk aversion lognormal version of the CCAPM.

The procedure of standard empirical tests will be followed to test the new model. Hence, conditional moments of the derived theoretical formulas are replaced by sample means, variances, and correlations in our study.

The representative agent maximizes his or her expected utility by

SOLUTION TO THE EQUITY PREMIUM PUZZLE

$$u(c_0) + \eta_t E_0 \left[ \sum_{t=0}^{\infty} \beta^{t+1} u(c_{t+1}) \right] \tag{15}$$

Here, $\eta_t$ is the sufficiency factor of the model that is determined at time t and is in the form of a coefficient. $E_0$ is the expectation operator conditional on the information available at the present time 0. β denotes the subjective time discount factor that is equal to 0.99 (Danthine & Donaldson, 2014); $u$ is an increasing, continuously differentiable concave utility function; and $c$ is per capita consumption.

Hence, the problem faced by a typical investor may be explained with the inclusion of the sufficiency factor of the model in Equation 9 as follows:

$$max_{\{\theta\}} u(c_t) + \beta \eta_t E_t [u(c_{t+1})]$$

$$\text{s.t.} \tag{16}$$

$$c_t = e_t - p_t \theta$$

$$c_{t+1} = e_{t+1} + d\theta$$

Here,

$c_t$ and $c_{t+1}$ = per capita consumption at times t and t+1, respectively;

$u$ = an increasing, continuously differentiable concave utility function;

$\beta$ = subjective time discount factor;

$\eta_t$ = sufficiency factor of the model that is determined at time t;

$e_t$ and $e_{t+1}$ = original consumption levels at times t and t+1, respectively, if the

investor did not buy any asset;

SOLUTION TO THE EQUITY PREMIUM PUZZLE

$p_t$                  = price of the asset at time t;

$d$                   = payoff that is equal to $[p_{t+1} + y_{t+1}]$ in equity

                        and 1 under risk-free asset; and

$\theta$                   = amount of assets.

The problem of the typical investor under dynamic optimization will be as follows:

$$max_{\{\theta_{t+1}, \ c_t\}} \{u(c_t) + \zeta_s E_t[\sum_{s=t}^{\infty} \beta^{s+1-t} u(c_{s+1})]\}$$

$$\text{s.t.} \tag{17}$$

$$\theta_{t+1} p_t + c_t \leq \theta_t y_t + \theta_t p_t$$

where $\zeta_s$ denotes the sufficiency factor of the model of equity investors. Furthermore, $c_t$ is a control variable, and $\theta_t$ is a state variable.

An alternative solution for the above problem of the typical investor by dynamic optimization is also provided in the alternative derivation of Equation 18 in the Appendix.

Since our forecasting abilities are limited and we are not able to see beyond the very near future, the sufficiency factor of the model is presumed a constant under the information available at present time. Hence, the sufficiency factor of the model is forecasted to be a constant coefficient.

When a decision regarding financial assets is made, the utility curve of the uncertain wealth of financial assets may shift upward or downward automatically according to the risk preferences of the investor. Hence, the sufficiency factor of the model is included in the problem of the typical investor because per capita consumption includes equity and risk-free assets.

The sufficiency factor of the model exists for investors investing in risk-free assets because it is possible for investors to sell risk-free assets before their maturity dates (i.e., investors may sell these assets to the Fed in the open market). As will be shown in subsequent pages, a no-trade equilibrium for the risk-free asset cannot exist. This leads to the possibility for the investor that the utility gained from the payoff that is worth 1 at the maturity date may be different from the utility gained from the uncertain payoff of the risk-free asset that is sold before the maturity date. For example, suppose that an investor investing in a risk-free asset bought the risk-free asset at \$95. The price of the asset increased to \$98 before its maturity date, and the investor decided to sell the asset at this price because of the macroeconomic conditions or the monetary policy of the Fed. The sufficiency factor of the model exists for the asset because $E_t[u(\$100)]$ (i.e., the predicted utility gained from certain face value at maturity) will be different from $E_t[u(\$98)]$ (i.e., the predicted utility gained from the uncertain \$98). Future uncertainty for the investor will emerge at the beginning of the period because of his or her probable decision that will occur before the maturity date.

The alternative solution of the problem of the typical investor by dynamic optimization has a binding constraint of $c_{t+1} = \theta_{t+1}y_{t+1} + \theta_{t+1}p_{t+1} - \theta_{t+2}p_{t+1}$ (see the alternative derivation of Equation 18 in Appendix). Hence, the per capita consumption at time t+1, $c_{t+1}$, is equal to $\theta_{t+1}y_{t+1} + \theta_{t+1}p_{t+1} - \theta_{t+2}p_{t+1}$, which is uncertain from the perspective of investors who compare the utilities of certain and uncertain wealth of financial assets at time t. Hence, the sufficiency factor of the model will exist for $E_t[u(c_{t+1})]$.

The economy is assumed to be frictionless. Moreover, one productive unit is assumed to produce output $y_t$ in period t, which is the period dividend. There is one equity share that is competitively traded. Hence, the following formulas hold true for the following assumptions and definitions of the new model:

SOLUTION TO THE EQUITY PREMIUM PUZZLE

(1) $u(c, \rho) = \frac{c^{1-\rho}}{1-\rho}$;

(2) $c$ denotes per capita consumption;

(3) $R_{e,t+1} = \frac{p_{t+1} + y_{t+1}}{p_t}$, where $p_{t+1}$ and $y_{t+1}$ are prices of the stock and dividends paid at

time t+1, respectively;

(4) $R_{f,t+1} = \frac{1}{q_t}$, where $q_t$ is the price of the risk-free asset;

(5) the growth rate of consumption, $x_{t+1} = \frac{c_{t+1}}{c_t}$, is log-normal;

(6) the growth rate of dividends, $z_{t+1} = \frac{y_{t+1}}{y_t}$, is log-normal; and

(7) $(x_t, z_t)$ are jointly lognormally distributed.

$$p_t u'(c_t) = \beta \zeta E_t [(p_{t+1} + y_{t+1}) u'(c_{t+1})] \quad \text{(fundamental pricing relationship)} \qquad (18)$$

See the derivation of Equation 18 in the Appendix for the proof.

$$ln\, E_t\, (R_{e,t+1}) \;\; = -\, ln\, \beta - ln\, \zeta + \rho \mu_x - \frac{1}{2} \rho^2 \sigma_x{}^2 + \rho \sigma_{x,z}\,, \qquad (19)$$

See the derivation of Equations 19 to 21 in the Appendix for the proof.

$$ln\, R_f = -\, ln\, \beta - ln\, \xi + \rho \mu_x - \frac{1}{2} \rho^2 \sigma_x{}^2\,, \qquad (20)$$

See the derivation of Equations 19 to 21 in the Appendix for the proof.

$$ln\, E(R_e) - ln\, R_f = ln\, \xi - ln\, \zeta \; + \rho \sigma_x{}^2\,, \qquad (21)$$

See the derivation of Equations 19 to 21 in the Appendix for the proof.

$$E(R_e) = \frac{E(x_{t+1})}{\beta \zeta E(x_{t+1}{}^{1-\rho})}\,, \qquad (22)$$

See the derivation of Equation 22 and Equation 23 in the Appendix for the proof.

SOLUTION TO THE EQUITY PREMIUM PUZZLE

$$ln\, E(R_e) = ln\, E(x_{t+1}) - ln\, \beta - ln\, \zeta\ - (1-\rho)\, \mu_x - \frac{1}{2}\, (1-\rho)^2 \sigma_x{}^2. \qquad (23)$$

See the derivation of Equation 22 and Equation 23 in the Appendix for the proof.

Here,

$\mu_x = E\,(ln\, x)$

$\sigma_x{}^2 = var\,(ln\, x)$

$\sigma_{x,z} = cov\,(ln\, x,\, ln\, z)$

$E\,(R_e) =$ mean return on equity for the period

$R_f =$ risk-free rate

$ln\, E\,(R_e) = ln$ of the mean return on equity for the period

$ln\, R_f = ln$ of the risk-free rate

$ln\, x =$ continuously compounded growth rate of consumption

$ln\, z =$ continuously compounded growth rate of dividends

$\beta =$ subjective time discount factor that is set as 0.99

$\zeta =$ sufficiency factor of the model for investors investing in equity

$\xi =$ sufficiency factor of the model for investors investing in risk-free assets.

For the entire economy to be in equilibrium, the following must hold true:

1- $\theta_t = \theta_{t+1} = \ldots = 1$ exists for the equity investor. This means that the representative agent possesses the entire equity share.

SOLUTION TO THE EQUITY PREMIUM PUZZLE

2- The possession of the entire equity share entitles the representative agent to all the economy's dividends, that is, $c_t = y_t$ (Danthine &Donaldson, 2014, p.277).

3- A no-trade equilibrium does not exist for the risk-free asset because the Fed is able to sell the risk-free asset to the representative agent or buy it in the open market from him or her.

The nonexistence of the no-trade equilibrium for the risk-free asset can also be observed mathematically. Since $u(c_t) = \frac{c_t^{1-\rho}}{1-\rho}$ (i.e., a strictly concave utility curve) is selected for the utility curve of the investors, $\lim_{c_t \to 0} u'(c_t) = \infty$ holds true. This selection ensures that it is never optimal for the investor to choose a zero-consumption level (Danthine &Donaldson, 2014, p.276).

When the fundamental equations of the CCAPM are derived, the constraint may alternatively be selected as $\theta_{t+1} p_t + c_t \le \theta_t y_t + \theta_t p_t$. Since the maximization of the objective function implies that the budget constraint will be binding, the constraint changes to $\theta_{t+1} p_t + c_t = \theta_t y_t + \theta_t p_t$.

If a no-trade equilibrium for the risk-free asset is assumed, $\theta_t = \theta_{t+1} = \ldots = 1$ holds true in the equilibrium. This implies that the equilibrium value of consumption is equal to zero from the budget constraint of risk-free assets, that is, $\theta_{t+1} q_t + c_t = \theta_t q_t$. Here, $q_t$ denotes the price of the risk-free asset at time t. This consumption value is obviously never optimal for the investor. Hence, $\theta_t = \theta_{t+1} = \ldots = 1$ does not hold true for the risk-free investor in the equilibrium.

4- With the inclusion of the sufficiency factor of the model of the investor, the equilibrium price must satisfy Equation 18.

Equations 18 to 21 were developed from Equations 10 to 13, respectively, by including the sufficiency factor of the model in Equations 10 to 13. Equation 23 is developed

SOLUTION TO THE EQUITY PREMIUM PUZZLE

from Equation 14 by including the sufficiency factor of the model of investors investing in equity.

The values of the sufficiency factor of the model for the investors investing in equity, the sufficiency factor of the model for the investors investing in risk-free assets, and the coefficient of relative risk aversion for the US economy for the period of 1889-1978 have been calculated using Equation 20, Equation 21, and Equation 23, respectively.

When the $\mu_x$ and $\sigma_x{}^2$ that are calculated according to Table 1 are substituted in Equation 20, Equation 21, and Equation 23, the following system of equations will be generated:

$$-ln\,\zeta - 0.017215(1-\rho) - \tfrac{1}{2}\,(1-\rho)^2 0.001250 = 0.039582, \tag{24}$$

$$-ln\,\xi + 0.017215\rho - 0.000625\rho^2 = -\,0.002082, \tag{25}$$

$$ln\,\xi - ln\,\zeta + 0.001250\rho = 0.059504. \tag{26}$$

[Insert Table 1 here] The solution of the above system of equations with the sum of squared errors (SSE) being $2.8\text{x}10^{-33}$ is:

$$\zeta \cong 0.961745$$

$$\xi \cong 1.019392$$

$$\rho \cong 1.033526.$$

The solution of the above system of equations was found using the NLSOLVE spreadsheet solver function of ExcelWorks LLC.

The subjective time discount factor is assigned a value of 0.99 for all calculations for the period. Input values of Equations 24 to 26 are taken with six decimal place accuracy

SOLUTION TO THE EQUITY PREMIUM PUZZLE

because the SSE of the solution of the above system of equations with the NLSOLVE spreadsheet solver function is the smallest at this accuracy.

### 3.3.2 Central Pricing Relationship of the CCAPM

We have

$$\beta \zeta_t E_t(\frac{u'(c_{t+1})}{u'(c_t)} R_{e,t+1}) = 1. \qquad (27)$$

The central pricing relationship of the CCAPM can be derived as follows by using Equation 27:

$$\xi_t R_{f,t+1} - \zeta_t E_t(R_{e,t+1}) = \beta \zeta_t \xi_t R_{f,t+1} \, cov_t(\frac{u'(c_{t+1})}{u'(c_t)}, R_{e,t+1}). \qquad (28)$$

See the derivation of Equation 28 in the Appendix for the proof.

## 4. Results and discussion

Many micro studies show us that the coefficient of relative risk aversion must be approximately equal to 1. Nevertheless, some economists believe that this coefficient can be as high as 2, 3 or 4. Despite this, there is a consensus among economists that the coefficient of relative risk aversion must be lower than 10.

The newly derived model sets the subjective time discount factor as 0.99 and assigns the coefficient of relative risk aversion a value of 1.033526. Since these values are compatible with empirical studies, they confirm that the model in this study provides the solution to the equity premium puzzle.

Since 1978 was automatically selected as the year for the determination of the behavior of investors toward risk due to the formulation of the problem, the solution of the above system of equations implies that as investors investing in risk-free assets increase their negative utility on an uncertain wealth value, those investing in equity decrease their negative utility on an uncertain wealth value. In other words, as investors investing in risk-free assets allocate negative utility for an uncertain wealth value, those investing in equity allocate positive utility for an uncertain wealth value, with the value of the coefficient of relative risk aversion being 1.033526.

Since the risk behavior of investors is exhibited with the help of a new tool, the definitions for risk-averse, risk-loving, and risk-neutral investors have also been reformulated; and some new definitions for the risk behavior of investors have been given.

Moreover, by assuming $c_t = \theta_t y_t + \theta_t p_t - \theta_{t+1} p_t$ for investors investing in equity and assuming $c_t = \theta_t q_t - \theta_{t+1} q_t$ for investors investing in risk-free assets from the budget constraint of Equation 17, Equations 1 to 4 can be used to detect the type of investor investing in 1977 and 1978.

This study certainly makes a significant contribution to the literature because the puzzle remained unsolved until now. Furthermore, a new tool, that is, the sufficiency factor of the model, has been used to classify investors as risk averse, risk loving, and risk neutral in this study.

## 5. Conclusions

This study provides the solution to the equity premium puzzle. The calculations of this new model show that the value of the coefficient of relative risk aversion is 1.033526 by assuming the value of the subjective time discount factor is 0.99. These values are found to be compatible with the empirical studies, confirming the validity of the derived model.

SOLUTION TO THE EQUITY PREMIUM PUZZLE

**Table 1**

*Summary Statistics for the U.S. Economy for the period of 1889-1978*

| Statistics | Value |
| --- | --- |
| Mean return on equity $E(R_e)$ | 1.0698 |
| Risk-free rate, $R_f$ | 1.008 |
| Mean growth rate of consumption, $E(x)$ | 1.018 |
| Standard deviation of the growth rate of consumption | 0.036 |
| Mean equity premium, $E(R_e) - R_f$ | 0.0618 |

Source: Adapted from *Handbook of the Equity Risk Premium* (pp. 19-20), by R. Mehra (2008). Copyright 2008 by Elsevier.

SOLUTION TO THE EQUITY PREMIUM PUZZLE

**Figure 1**

*Certain and uncertain utility curves of the risk-averse investor with the sufficiency factor of the model*

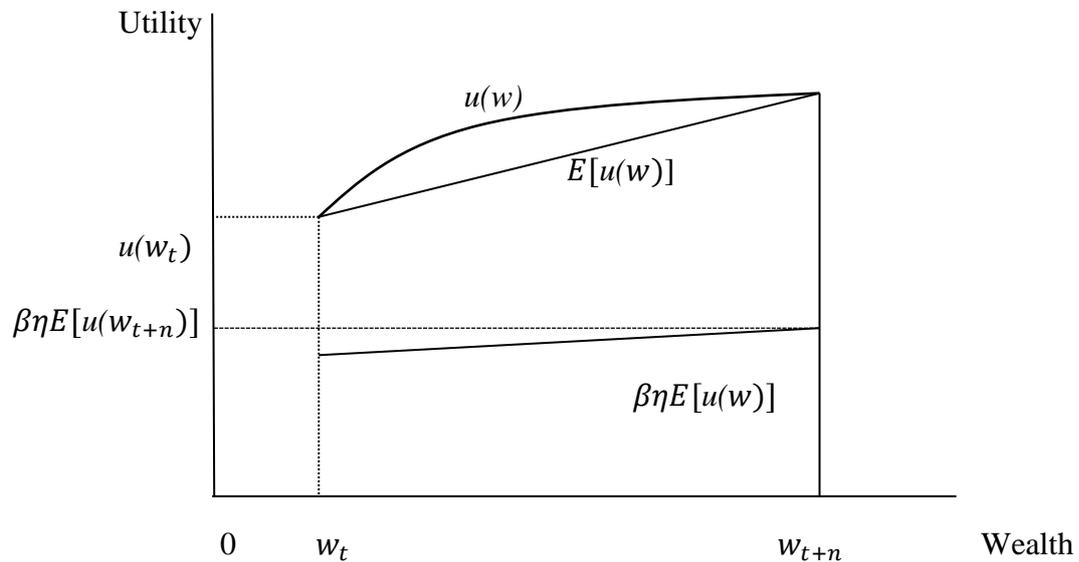

*Note.* The figure demonstrates the certain utility curve *u(w)*, the uncertain utility curve $E[u(w)]$ and the uncertain utility curve $\beta\eta E[u(w)]$ that includes the sufficiency factor of the model $(\eta)$ and the subjective time discount factor $(\beta)$ of a risk-averse investor who allocates negative utility for the uncertain wealth values.

**Figure 2**

*The utility curves of the investor with ρ and η*

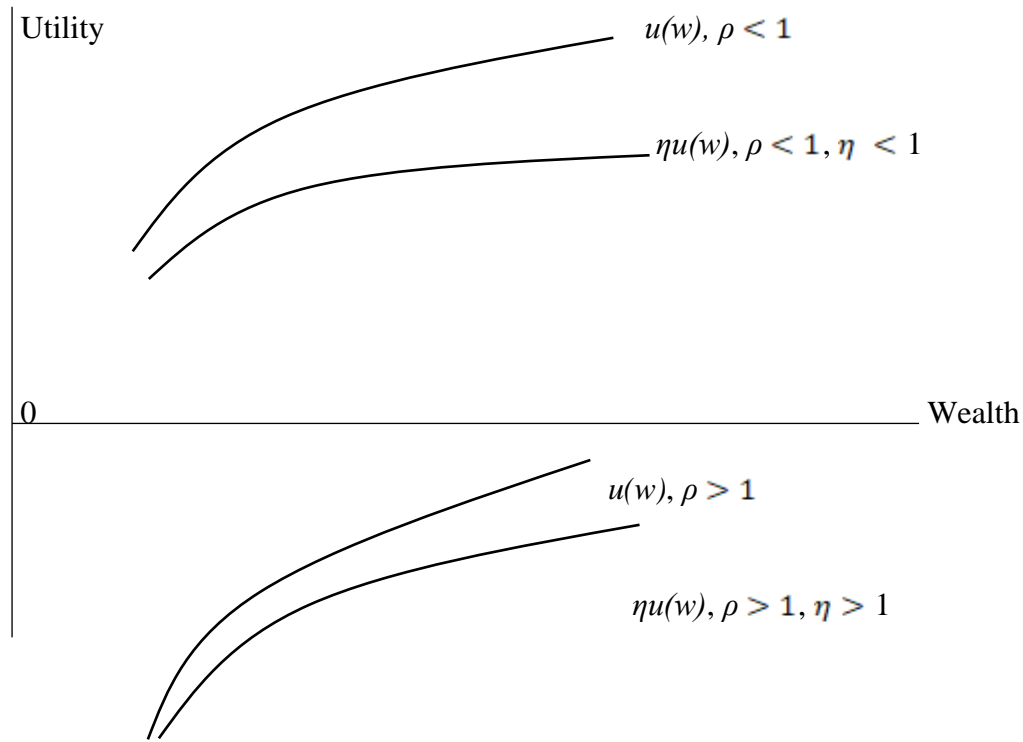

*Note.* The figure demonstrates the utility curves of an investor who allocates negative utility

for the uncertain wealth values with ρ and η, where η is the sufficiency factor of the model and ρ is the coefficient of relative risk aversion.

**Figure 3**

*The utility curves of the investor with ρ and η*

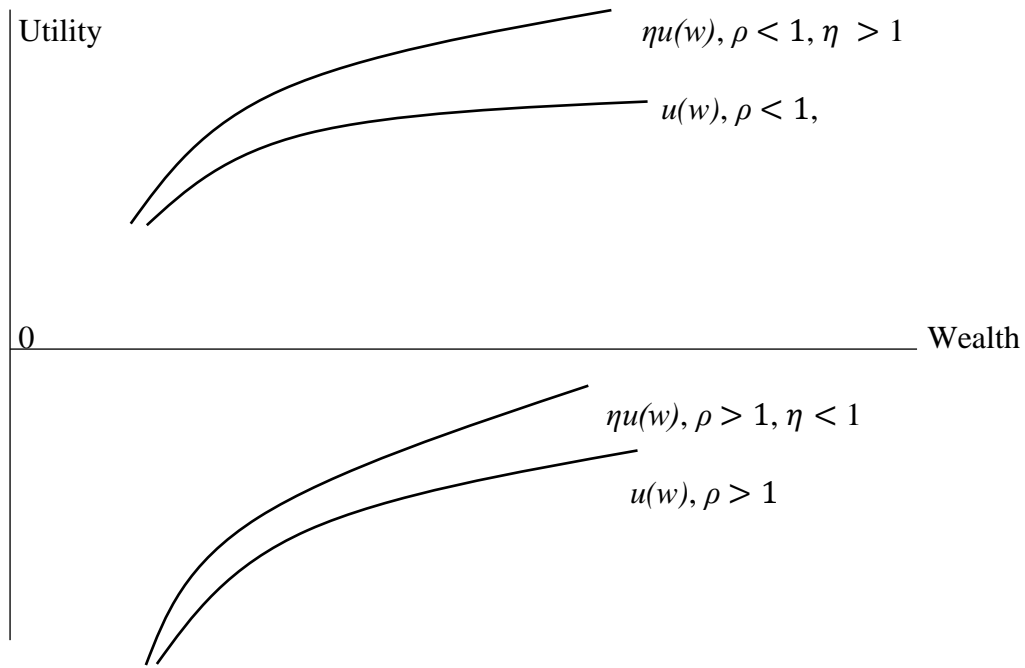

*Note.* The figure demonstrates the utility curves of an investor who allocates positive utility for the uncertain wealth values with $\rho$ and $\eta$, where $\eta$ is the sufficiency factor of the model and $\rho$ is the coefficient of relative risk aversion.

**Appendix**

**Derivation of Equation 18** (Cochrane, 2009, p. 5)

Now, we place the constraints in the objective function and set the derivative with respect to $\theta$ equal to zero in Equation 16 with $d = p_{t+1} + y_{t+1}$. This results in Equation 18.

**Alternative Derivation of Equation 18 by Dynamic Optimization**

The problem of the typical investor will be

$$max_{\{\theta_{t+1}, \ c_t\}} \{u(\ c_t) + \zeta E_t[\sum_{s=t}^{\infty} \beta^{s+1-t} u(c_{s+1})]\}$$

s.t.                 (A.1)

$$\theta_{t+1} p_t \ + \ c_t \ \leq \ \theta_t y_t + \theta_t p_t$$

where $\zeta$ denotes the sufficiency factor of the model of the equity investors. Furthermore, $c_t$ is a control variable, and $\theta_t$ is a state variable. The right- and left-hand side of the budget constraint denote total resources and use of those resources, respectively.

We will use Bellman's optimality principle to solve the problem. Bellman's optimality principle allows us to write the problem in A.1 as the following two-period problem:

$$V_t(\theta_t) = max_{\{\theta_{t+1}, \ c_t\}} \{u(\ c_t) + \beta \zeta E_t[\ V_{t+1}(\theta_{t+1})]\}$$     (A.2)

s.t.

$$\theta_{t+1} = \frac{(y_t + p_t)\theta_t - c_t}{p_t} \ .$$     (A.3)

We will write A.2 and A.3 in Bellman form to derive the first-order conditions by using the Lagrangian as follows:

SOLUTION TO THE EQUITY PREMIUM PUZZLE

$$V_t(\theta_t) = max_{\{\theta_{t+1},\, c_t\}}\, [u(c_t) + \beta\zeta E_t(V_{t+1}(\theta_{t+1}))] + \lambda_t[\theta_{t+1}p_t\ +\ c_t - \theta_t y_t - \theta_t p_t\,].$$

(A.4)

Then, we differentiate A.4 with respect to $c_t$ and $\theta_{t+1}$ to obtain the first-order conditions:

$$u'(c_t) + \lambda_t = 0,$$

(A.5)

$$\beta\zeta E_t\,[\,V'_{t+1}\,(\theta_{t+1})] + \lambda_t p_t = 0.$$

(A.6)

Hence, from A.5 and A.6,

$$u'(c_t) = \zeta\beta E_t\,[\,V'_{t+1}\,(\theta_{t+1})]\frac{1}{p_t}.$$

(A.7)

To get the envelope condition, just take the derivative of A.4 with respect to $\theta_t$:

$$V'_t(\theta_t) = -\lambda_t\,(y_t + p_t).$$

(A.8)

After shifting up one period, we obtain

$$V'_{t+1}(\theta_{t+1}) = -\lambda_{t+1}\,(y_{t+1} + p_{t+1})$$

(A.9)

and

$$-\,\lambda_{t+1} = u'(c_{t+1})$$

(A.10)

holds true according to A.5.

We substitute A.10 in A.9 to obtain

$$V'_{t+1}(\theta_{t+1}) = u'(c_{t+1})\,(y_{t+1} + p_{t+1}).$$

(A.11)

We substitute A.11 in A.7 to obtain

SOLUTION TO THE EQUITY PREMIUM PUZZLE

$$p_t u'(c_t) = \zeta \beta E_t[u'(c_{t+1})(p_{t+1} + y_{t+1})]. \tag{A.12}$$

**Derivation of Equations 19 to 21** (Mehra, 2008, pp. 17-19)

Since the equity price is homogeneous of degree 1 in y, the equity price is in the form of

$$p_t = v y_t, \tag{A.13}$$

where $v$ is a constant coefficient.

By the fundamental pricing relationship, we have

$$v y_t = \beta \zeta E_t[(v y_{t+1} + y_{t+1}) \frac{u'(c_{t+1})}{u'(c_t)}]. \tag{A.14}$$

Hence,

$$v = \frac{\beta \zeta \, E_t(z_{t+1} x_{t+1}^{-\rho})}{1 - \beta \, \zeta E_t(z_{t+1} x_{t+1}^{-\rho})}. \tag{A.15}$$

By definition,

$$R_{e,t+1} = \frac{p_{t+1} + y_{t+1}}{p_t} \tag{A.16}$$

or

$$R_{e,t+1} = \left(\frac{v+1}{v}\right)\left(\frac{y_{t+1}}{y_t}\right) = \left(\frac{v+1}{v}\right) z_{t+1}. \tag{A.17}$$

Taking the conditional expectation on both sides of Equation A.17 results in

$$E_t(R_{e,t+1}) = \left(\frac{v+1}{v}\right) E_t(z_{t+1}). \tag{A.18}$$

Substituting

SOLUTION TO THE EQUITY PREMIUM PUZZLE

$$\left(\frac{v+1}{v}\right) = \frac{1}{\beta\,\zeta E_t(z_{t+1}x_{t+1}^{-\rho})} \tag{A.19}$$

in $E_t(R_{e,t+1})$ results in

$$\frac{E_t(z_{t+1})}{\beta\zeta E_t(z_{t+1}x_{t+1}^{-\rho})}. \tag{A.20}$$

Similarly,

$$R_{f,t+1} = \frac{1}{\beta\xi E_t(x_{t+1}^{-\rho})}. \tag{A.21}$$

By using the lognormal properties and the fact that conditional expectations are replaced with sample means in standard empirical tests, we have

$$E(R_e) = \frac{exp(\mu_z + \frac{1}{2}\sigma_z^2)}{\beta\zeta exp\left[\mu_z - \rho\mu_x + \frac{1}{2}(\sigma_z^2 + \rho^2\sigma_x^2 - 2\,\rho\sigma_{x,z})\right]} \tag{A.22}$$

and

$$R_f = \frac{1}{\beta\xi exp\left(-\rho\mu_x + \frac{1}{2}\rho^2\sigma_x^2\right)}. \tag{A.23}$$

If we take ln on both sides, we obtain

$$ln\,E(R_e)\;\; = -\,ln\,\beta - ln\,\zeta\, + \rho\mu_x - \frac{1}{2}\rho^2\sigma_x^2 + \rho\sigma_{x,z} \tag{A.24}$$

and

$$ln\,R_f\;\; = -\,ln\,\beta - ln\,\xi + \rho\mu_x - \frac{1}{2}\rho^2\sigma_x^2. \tag{A.25}$$

Subtracting $ln\,R_f$ from $ln\,E\,(R_e)$ results in

$$ln\,E\,(R_e) - ln\,R_f = ln\,\xi - ln\,\zeta\, + \rho\sigma_{x,z}. \tag{A.26}$$

Since the equilibrium condition sets $x = z$, we have

$$ln\, E\,(R_e) - ln\, R_f = ln\, \xi - ln\, \zeta + \rho\sigma_x{}^2. \tag{A.27}$$

**Derivation of Equation 22 and Equation 23** (Danthine & Donaldson, 2014, pp. 289-291)

The equity price is of the form

$$p_t = vy_t, \tag{A.28}$$

where $v$ is a constant coefficient.

From Equation 18, we have

$$vy_t = \beta\zeta E_t[(vy_{t+1} + y_{t+1})\frac{u'(c_{t+1})}{u'(c_t)}]. \tag{A.29}$$

If conditional expectations are replaced with sample means according to the procedures of standard empirical tests, we obtain

$$v = \beta\zeta E\,[(v+1)\,\frac{y_{t+1}}{y_t}\,x_{t+1}{}^{-\rho}]. \tag{A.30}$$

The market clearing condition requires that

$$\frac{y_{t+1}}{y_t} = x_{t+1}. \tag{A.31}$$

Therefore,

$$v = \beta\zeta E\,[(v+1)\,x_{t+1}{}^{1-\rho}] \tag{A.32}$$

or

$$v = \frac{\beta\zeta E(x_{t+1}{}^{1-\rho})}{1-\beta\zeta E(x_{t+1}{}^{1-\rho})}. \tag{A.33}$$

SOLUTION TO THE EQUITY PREMIUM PUZZLE

Hence, $v$ is indeed a constant.

Since

$$R_{e,t+1} = \frac{p_{t+1}+y_{t+1}}{p_t} = \frac{v+1}{v}\frac{y_{t+1}}{y_t} = \frac{v+1}{v}x_{t+1}, \qquad (A.34)$$

the following holds true if expectations are taken on both sides of Equation A.34 and conditional expectations are replaced with sample means according to the procedures of standard empirical tests:

$$E_t(R_{e,t+1}) = E(R_e) = \frac{v+1}{v}E(x_{t+1}) = \frac{E(x_{t+1})}{\beta\zeta E(x_{t+1}^{1-\rho})} \qquad (A.35)$$

or

$$E_t(R_{e,t+1}) = E(R_e) = \frac{E(x_{t+1})}{\beta\zeta exp\left[(1-\rho)\,\mu_x + \frac{1}{2}(1-\rho)^2\sigma_x^2\right]}. \qquad (A.36)$$

Taking the ln of both sides results in

$$ln\,E(R_e) = ln\,E(x_{t+1}) - ln\,\beta - ln\,\zeta - (1-\rho)\,\mu_x - \frac{1}{2}(1-\rho)^2\sigma_x^2. \qquad (A.37)$$

**Derivation of Equation 28**

We have

$$\beta\zeta_t E_t(\frac{u'(c_{t+1})}{u'(c_t)}R_{e,t+1}) = 1. \qquad (A.38)$$

By using $E(X.Y) = E(X) \cdot E(Y) + cov\,(X,Y)$, Equation A.38 can be written as

$$\beta\zeta_t E_t(\frac{u'(c_{t+1})}{u'(c_t)})E_t(R_{e,t+1}) + \beta\zeta_t cov_t(\frac{u'(c_{t+1})}{u'(c_t)}, R_{e,t+1}) = 1. \qquad (A.39)$$

We have

SOLUTION TO THE EQUITY PREMIUM PUZZLE

$$R_{f,t+1} = \frac{1}{\beta \xi_t E_t(\frac{u'(c_{t+1})}{u'(c_t)})} \qquad (A.40)$$

or

$$E_t\left(\frac{u'(c_{t+1})}{u'(c_t)}\right) = \frac{1}{\beta \xi_t R_{f,t+1}} \; . \qquad (A.41)$$

Substituting Equation A.41 in Equation A.39 results in

$$\frac{\zeta_t}{\xi_t R_{f,t+1}} . E_t(R_{e,t+1}) + \beta \zeta_t cov_t\left(\frac{u'(c_{t+1})}{u'(c_t)}, R_{e,t+1}\right) = 1. \qquad (A.42)$$

After algebraic operations, we possess

$$\xi_t R_{f,t+1} - \zeta_t E_t(R_{e,t+1}) = \beta \zeta_t \xi_t R_{f,t+1} \, cov_t\left(\frac{u'(c_{t+1})}{u'(c_t)}, R_{e,t+1}\right). \qquad (A.43)$$